\documentclass{article}
\usepackage{amsmath}
\usepackage{graphicx}
\usepackage{latexsym}

\title{Strings in the Einstein's paradigm of matter}
\author{Vladimir Dzhunushaliev
\thanks{
E-mail: dzhun@hotmail.kg}}
\begin{document}
\maketitle

\begin{center}
\textit{
Institut f\"ur Theoretische Physik,  
Freie Universit\"at Berlin, Arnimallee 14, D-14195, Berlin, Germany\\
and \\
Dept. Phys. and Microel. Engineer., Kyrgyz-Russian
Slavic University\\
Bishkek, Kievskaya Str. 44, 720000, Kyrgyz
Republic}
\end{center}

\begin{abstract}
Spherically symmetric solutions with a flux of electric and magnetic 
fields in Kaluza-Klein gravity are considered. It is shown that under the 
condition (electric charge $q$) $\approx$ (magnetic charge $Q$) ($q>Q$) 
these solutions are like a flux tube stretched between two Universes. 
The longitudinal size of this tube depends on the value of 
$\delta = 1 - Q/q$. The cross section of the tube can be chosen 
$\approx l_{Pl}$. In this case this flux tube looks like a 1-dimensional 
object (thread) stretched between two Universes. The propagation of 
gravitational waves on the thread is considered. The corresponding 
equations are very close to the classical string equations. This 
result allows us to say that the thread between two Universes is 
similar to the string attached to two $D-$branes. 
\end{abstract}

\section{Introduction}

In string theory there is a problem which can not be resolved
inside of the theory : Is there an inner structure of the string ? If
the string is some kind of matter then string theory can not be
a fundamental theory of Everything. We need to have a theory for
the string matter. If such matter has an elementary structure then
we must have a theory for this structure and so on up to infinity.
In fact this is analogous to the well known old problem
of the inner structure of electron. Probably
the most radical resolution of this problem was suggested
by Einstein. As usually his approach is very revolutionary :
the electron does not have any material structure, it is a bridge
between two asymptotically flat spaces (a wormhole in the modern
language). One can say that this is Einstein's paradigm for the 
fundamental structure of matter. 
\par
Surprisingly, this point of view is more adaptable to strings then
to point particles. The reason is that it is very difficult to find
vacuum wormhole solutions of Einstein equations, but string-like solutions
arise as very natural objects in \emph{vacuum 5D Kaluza-Klein theory}.
In this paper we would like to present such solutions and interpret them 
as string-like objects. 

\section{String-like solutions}

In common case the 5D metric $G_{AB}$ ($A,B = 0,1,2,3,5$) has the 
following form 
\begin{equation}
G_{AB} = \left (
\begin{array}{ll}
g_{\mu \nu} - \phi ^2 A_\mu A_\nu & -\phi ^2 A_\mu \\
-\phi ^2 A_\mu                    & -\phi ^2
\end{array}
\right )
\label{sec1-1a}
\end{equation}
$\mu , \nu = 0,1,2,3$ are the 4D indices. According to the Kaluza - Klein 
point of view $g_{\mu \nu}$ is the 4D metric; 
$A_\mu$ is the usual electromagnetic potential and $\phi$ is some scalar 
field. The 5D Einstein vacuum equations are 
\begin{equation}
R_{AB} - \frac{1}{2}G_{AB} R = 0,
\label{sec1-2}
\end{equation}
where $R_{AB}$ is the 5D Ricci tensor 
and after the dimensional reduction we have the following 4D equations 
for the (electrogravity + scalar field) theory (for the reference see, 
for example, \cite{wesson}) 
\begin{eqnarray}
R_{\mu \nu} - \frac{1}{2} g_{\mu \nu} R & = & 
-\frac{\phi^2}{2} T_{\mu \nu} - \frac{1}{\phi} 
\left [
\nabla _\mu \left (
            \partial _\nu \phi
            \right ) - g_{\mu \nu} \Box \phi 
\right ] ,
\label{sec1-2a}\\
\nabla _\nu F^{\mu \nu} & = & -3 \frac{\partial _\nu \phi}{\phi} F^{\mu \nu} ,
\label{sec1-2b}\\
\Box \phi & = & - \frac{\phi^3}{4} F^{\alpha \beta} F_{\alpha \beta} .
\label{sec1-2c}
\end{eqnarray}
where $R_{\mu \nu}$ is the 4D Ricci tensor; 
$F_{\mu \nu} = \partial_\mu A_\nu - \partial_\nu A_\mu$ is the 
4D Maxwell tensor and $T_{\mu \nu}$ is the energy-momentum tensor 
for the electromagnetic field. 
\par 
Now we would like to present the spherically symmetric solutions with
nonzero flux of electric and/or magnetic fields \cite{dzhsin}.
For our spherically symmetric 5D metric we take
\begin{eqnarray}
ds^2 & = & \frac{dt^{2}}{\Delta(r)} - l_0^2 \Delta(r) e^{2\psi (r)}
\left [d\chi +  \omega (r)dt + Q \cos \theta d\varphi \right ]^2
\nonumber \\
&-& dr^{2} - a(r)(d\theta ^{2} +
\sin ^{2}\theta  d\varphi ^2),
\label{sec1-1}
\end{eqnarray}
where $\chi $ is the 5$^{th}$ extra coordinate;
$r,\theta ,\varphi$ are $3D$  spherical-polar coordinates;
$n$ is constant; $r \in \{ -r_H , +r_H \}$
($r_H$ may be equal to $\infty$), $l_0$ is some constant; $Q$ is 
the magnetic charge as $(\theta \varphi)$ -component of the Maxwell tensor 
$F_{23} = -\sin\theta$ and consequently the radial 
magnetic field is $H_r = Q/a$.
\par 
In our case  the metric \eqref{sec1-1} give us the 4D metric 
\begin{equation}
d \stackrel{(4)}{s^2} = \frac{dt^{2}}{\Delta(r)} - 
dr^{2} - a(r)(d\theta ^{2} + \sin ^{2}\theta  d\varphi ^2),
\label{sec1-2d}
\end{equation}
and the following components of electromagnetic potential 
$A_\mu$ 
\begin{equation}
A_0 = \omega (r) \quad 
\text{and} \quad 
A_\varphi = Q \cos \theta
\label{sec1-2e}
\end{equation}
that gives the Maxwell tensor 
\begin{equation}
F_{10} = \omega (r) ' \quad 
\text{and} \quad 
F_{23} = -Q \sin \theta .
\label{sec1-2f}
\end{equation}
This means that we have radial Kaluza-Klein 
electric $E_r \propto F_{01}$ and magnetic 
$H_r \propto F_{23}$ fields.
\par
Substituting this ansatz into the 5D Einstein vacuum equations
\eqref{sec1-2} gives us 
\begin{eqnarray}
\frac{\Delta ''}{\Delta} - \frac{{\Delta '}^2}{\Delta^2} + 
\frac{\Delta 'a'}{\Delta a} + \frac{\Delta ' \psi '}{\Delta} + 
\frac{q^2}{a^2 \Delta ^2}e^{-4 \psi} & = & 0,
\label{sec1-3}\\
\frac{a''}{a} + \frac{a'\psi '}{a} - \frac{2}{a} +
\frac{Q^2}{a^2} \Delta e^{2\psi} & = & 0,
\label{sec1-5}\\
\psi '' + {\psi '}^2 + \frac{a'\psi '}{a} -
\frac{Q^2}{2a^2} \Delta e^{2\psi} & = & 0,
\label{sec1-6}\\
- \frac{{\Delta '}^2}{\Delta^2} + \frac{{a'}^2}{a^2} - 
2 \frac{\Delta ' \psi '}{\Delta} - \frac{4}{a} + 
4 \frac{a' \psi '}{a} + 
\frac{q^2}{a^2 \Delta ^2} e^{-4 \psi} + 
\frac{Q^2}{a^2} \Delta e^{2\psi} & = & 0
\label{sec1-7}
\end{eqnarray}
$q$ is some constant which physical sense will be discussed below; 
these equations are derived after substitution the expression 
\eqref{sec1-7e} for the electric field 
in the initial Einstein's equatons \eqref{sec1-2}. 
The 5D $(\chi t)$-Einstein's equation (4D Maxwell equation) is taken as 
having the following solution 
\begin{equation}
  \omega ' = \frac{q}{l_0 a \Delta ^2} e^{-3 \psi} .
\label{sec1-7e}
\end{equation}
For the determination of the physical sense of the constant $q$ let us 
write the $(\chi t)$-Einstein's equation in the following way :
\begin{equation}
\left( l_0 \omega ' \Delta ^2 e^{3\psi} 4 \pi a \right)' = 0.
\label{sec1-7a}
\end{equation}
The 5D Kaluza - Klein gravity after the dimensional reduction says us that 
the Maxwell tensor is 
\begin{equation}
F_{\mu \nu} = \partial_\mu A_\nu - \partial _\nu A_\mu . 
\label{sec1-7e1}
\end{equation}
That allows us to write in our case the electric field 
as $E_r = \omega '$. 
Eq.\eqref{sec1-7a} with the electric field defined by \eqref{sec1-7e1} 
can be compared with the Maxwell's equations in a continuous medium 
\begin{equation}
\mathrm{div} \mathcal {\vec D} = 0
\label{sec1-7e2}
\end{equation}
where $\mathcal {\vec D} = \epsilon \vec E$ is an electric displacement 
and $\epsilon$ is a dielectric permeability. 
Comparing Eq. \eqref{sec1-7a}
with Eq. \eqref{sec1-7e2} we can say that the magnitude 
$q/a = \omega ' \Delta^2 e^{3\psi}$ is like to the electric displacement 
and in this case the dielectric permeability is $\epsilon = \Delta^2 e^{3\psi}$. 
It means that $q$ can be taken as the Kaluza-Klein electric charge 
because the flux of electric field is 
$\mathbf{\Phi} = 4\pi a\mathcal D = 4\pi q$.
\par 
Eq. \eqref{sec1-7} gives us the
following relationship between the Kaluza-Klein electric 
and magnetic charges
\begin{equation}
\label{sec1-7d}
1 = \frac{q^2 + Q^2}{4a(0)}
\end{equation}
where $a(0) = a(r=0)$. 
From Eq. \eqref{sec1-7d} it is seen that the charges can be parameterized
as $q = 2 \sqrt{a(0)}\sin \alpha$ and $Q = 2 \sqrt{a(0)}\cos \alpha$.
\par
As the relative strengths of the Kaluza-Klein fields are
varied it was found \cite{dzhsin} that the solutions to the metric in
Eq. \eqref{sec1-1} evolve in the following way :
\begin{enumerate} 
\item 
$0 \leq Q < q$. The solution is \emph{a regular flux tube}. 
The throat between the surfaces at $\pm r_H$ is filled with both 
electric and magnetic fields. The longitudinal
distance between the $\pm r_H$ surfaces increases, and
the cross-sectional size does not increase as rapidly
as $r \rightarrow r_H$ with $q \rightarrow Q$. Essentially,
as the magnetic charge is increased one can think that
the $\pm r_H$ surfaces are taken to $\pm \infty$ and 
the cross section becomes constant. The radius $r=r_H$ is defined by the 
following way $ds^2(r = \pm r_H) = 0$. 
\item 
$Q = q$. In this case the solution is \emph{an infinite flux tube} filled
with constant electric and magnetic fields. The cross-sectional
size of this solution is constant ($ a= const.$). A similar object was
derived within the context of 4D dilaton gravity in Ref. \cite{davidson}. 
\item 
$0 \leq q < Q$. In this case we have \emph{a singular flux tube} located 
between two (+) and (-) electric and magnetic 
charges located at $\pm r_{sing}$. Thus the longitudinal 
size of this object is again finite, but now the cross 
sectional size decreases as $r \rightarrow r_{sing}$. At 
$r = \pm r_{sing}$ this solution has real singularities which 
we interpret as the locations of the charges. 
\end{enumerate} 
\par
\begin{figure}
\begin{center}
\fbox{
\includegraphics[height=5cm,width=5cm]{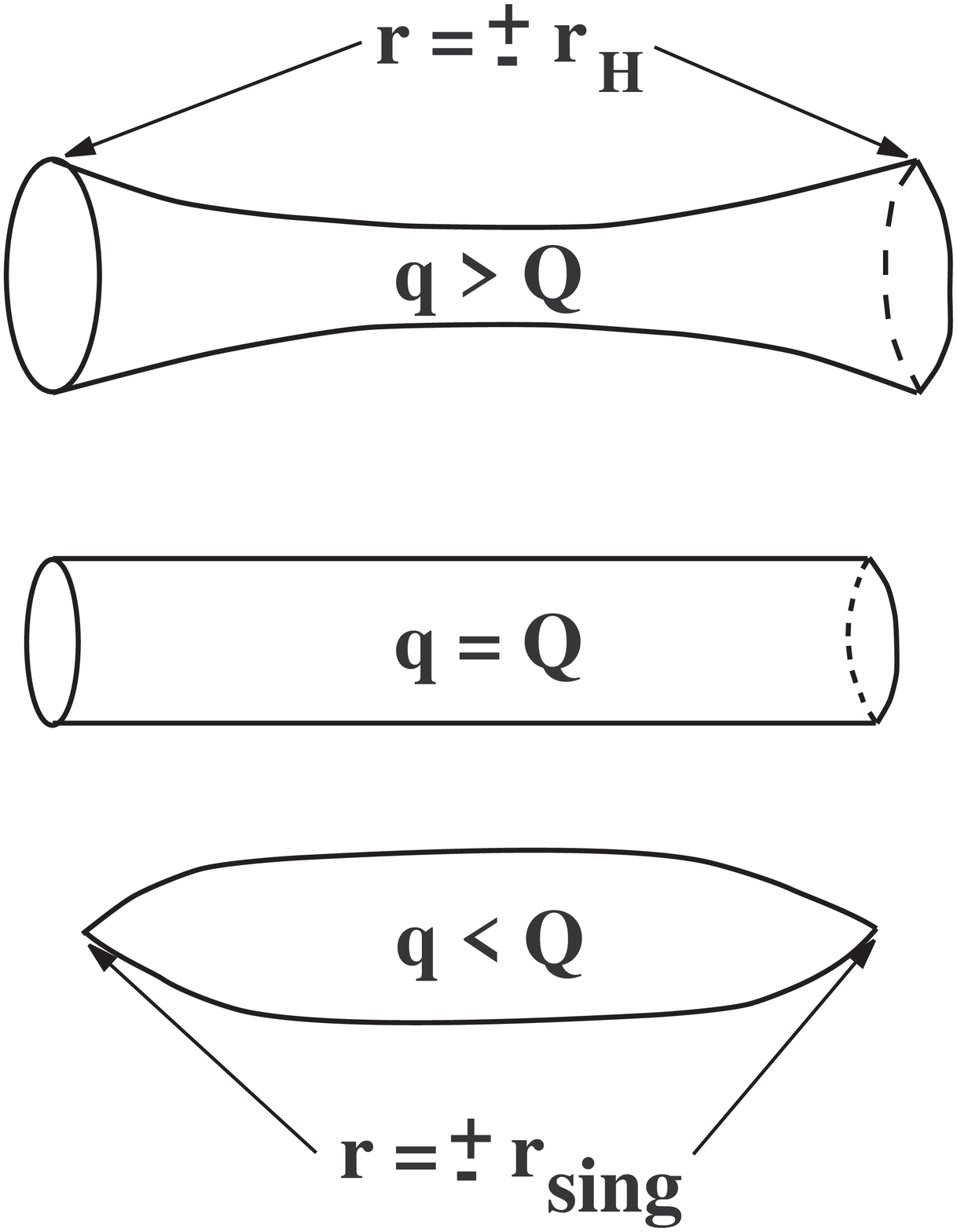}}
\caption{The flux tubes with different relation between 
electric $q$ and magnetic $Q$ charges.}
\label{fig4}
\end{center}
\end{figure}

 
The evolution of the solution from a regular flux tube, to
an infinite flux tube, to a singular flux tube, as the
relative magnitude of the charges is varied, is presented in
Fig.\eqref{fig4}. 
\par
Most important for us here is the case with $q \approx Q$
(but $q > Q$). From Eq. \eqref{sec1-7d} we have 
$q = 2 a_0 \sin (\pi /4 + \delta) \approx a_0 \sqrt{2} (1 + \delta)$ 
and $Q \approx a_0 \sqrt{2} (1 - \delta)$. 

\section{Long flux tube solutions}

Now we would like to consider the case with $q \approx Q$ 
($q > Q$) ($|r| \leq r$). As we see above the case $q = Q$ is the 
infinite flux tube 
with (electric field) = (magnetic field). Our basic aim is to show 
that \emph{every such object with $2 \delta = 1 - Q/q \ll 1$ and cross 
section $\approx l^2_{Pl}$ can be considered as a string-like object}. 
\par
\begin{figure}
\begin{center}
\fbox{
\includegraphics[height=5cm,width=5cm]{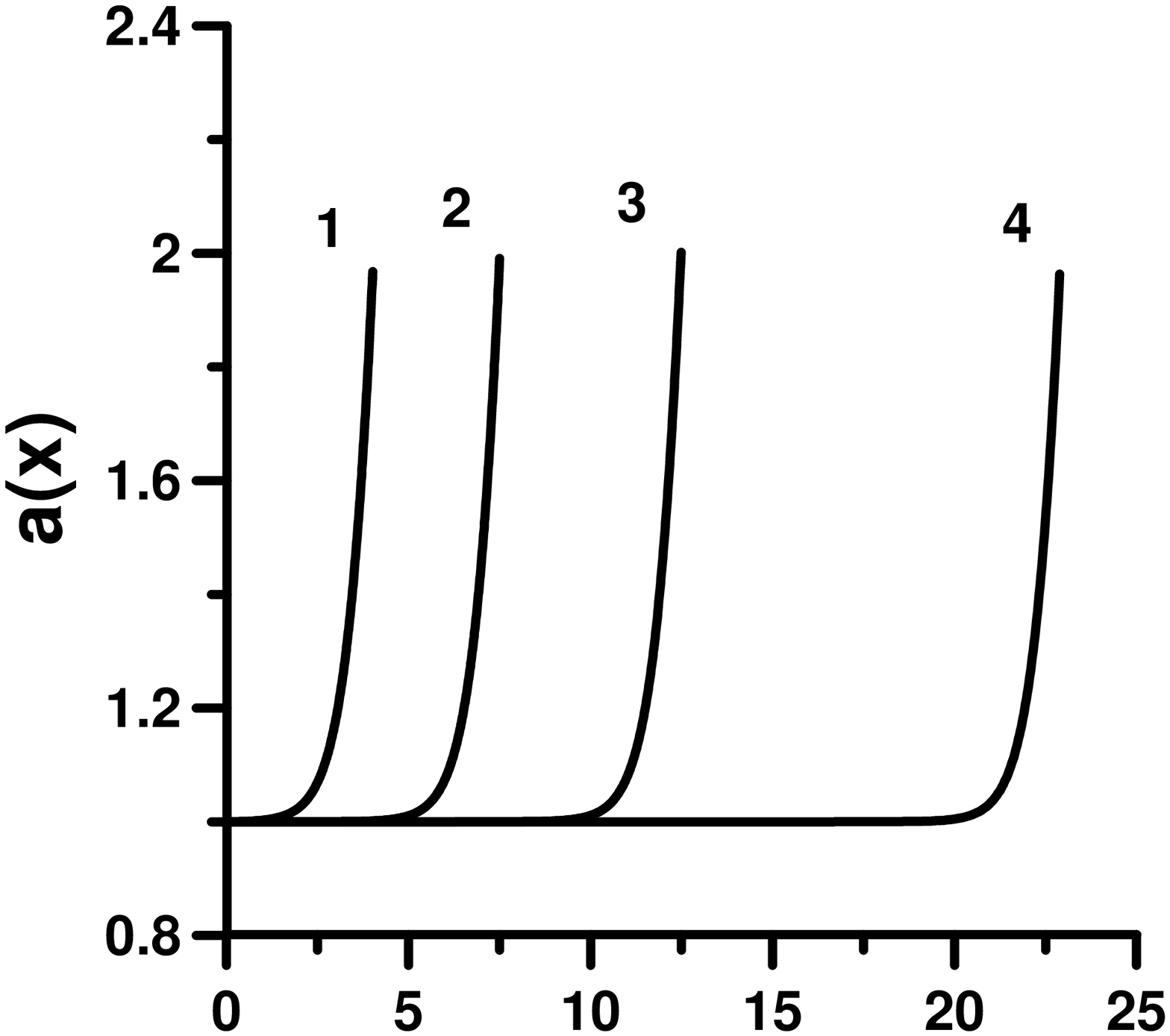}}
\caption{The functions $a(x)$ with the different magnitudes of $\delta$.
The curves 1,2,3,4 correspond to the following magnitudes of 
$\delta = 10^{-3}, 10^{-6}, 10^{-10}, 10^{-13}$.}
\label{fig1}
\fbox{
\includegraphics[height=5cm,width=5cm]{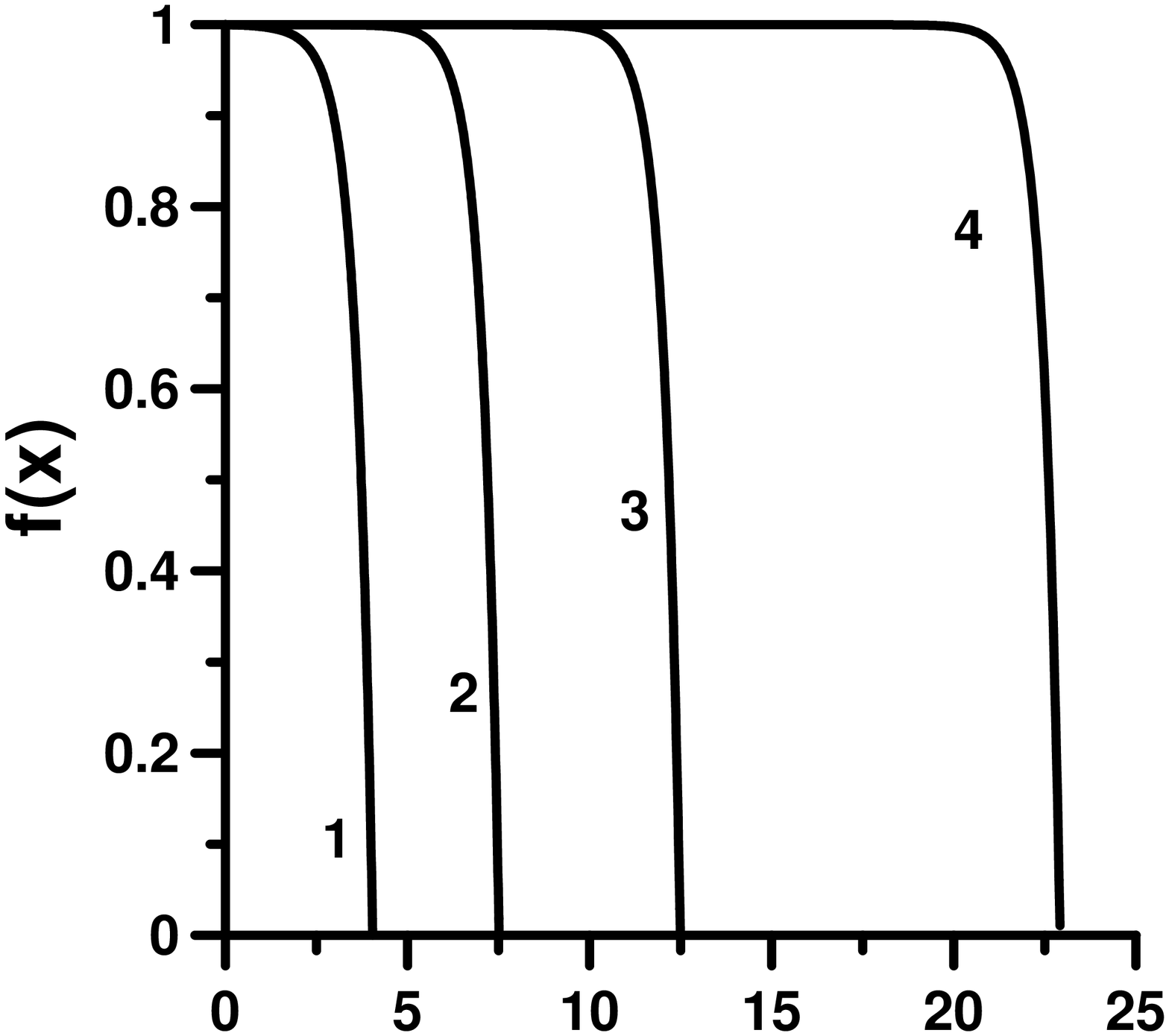}}
\caption{The functions $f(x)$ with the different magnitudes of $\delta$.
The curves 1,2,3,4 correspond to the following magnitudes of 
$\delta = 10^{-3}, 10^{-6}, 10^{-10}, 10^{-13}$.}
\label{fig2}

\fbox{
\includegraphics[height=5cm,width=5cm]{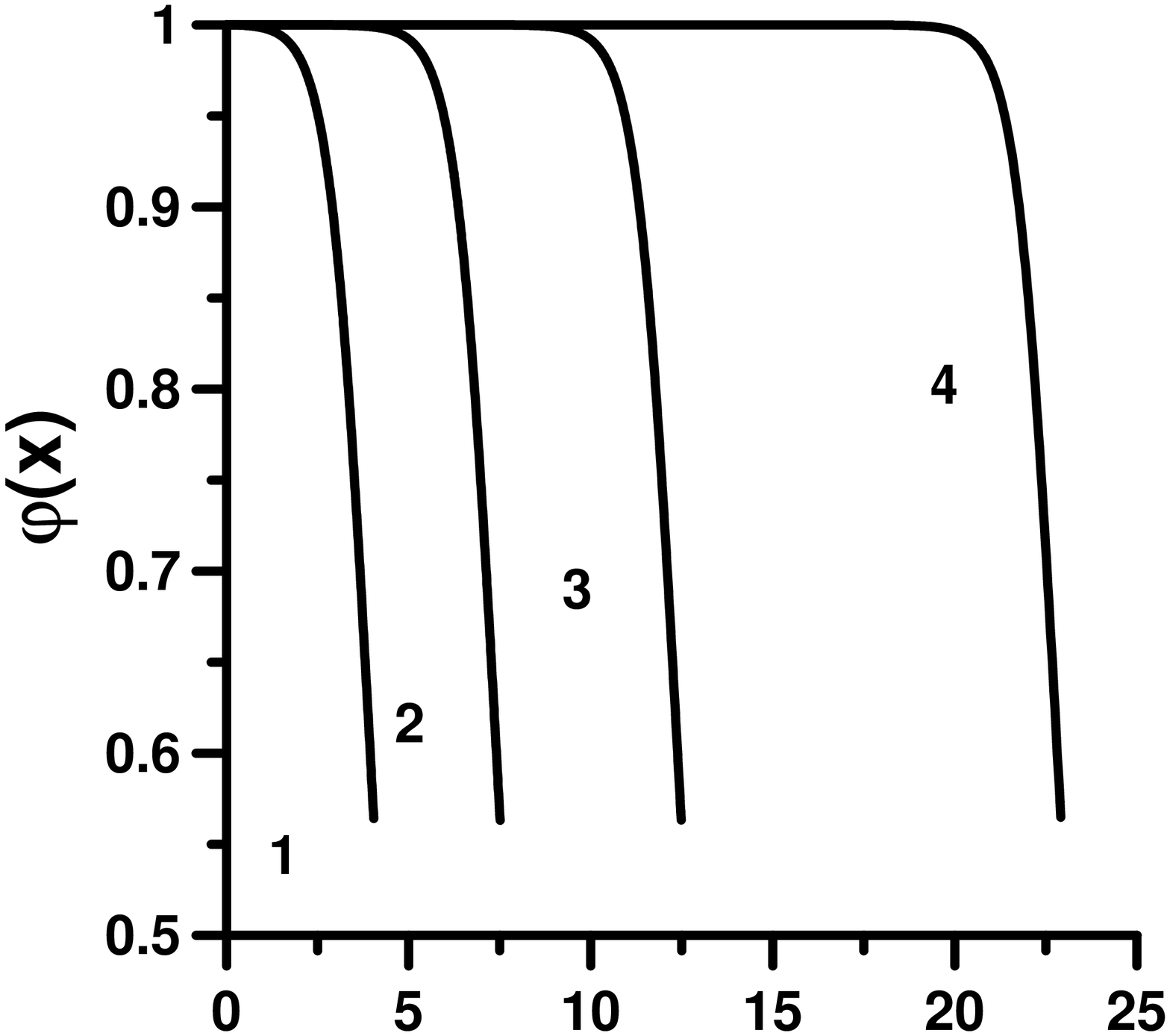}}
\caption{The functions $\phi(x)$ with the different magnitudes of $\delta$.
The curves 1,2,3,4 correspond to the following magnitudes of 
$\delta = 10^{-3}, 10^{-6}, 10^{-10}, 10^{-13}$.}
\label{fig3}
\end{center}
\end{figure}

 
To do this we will numerically investigate solutions of Eq's 
\eqref{sec1-3}-\eqref{sec1-7} for different $\delta$'s. 
The numerical investigations of our system of equations is complicated 
as it is very insensitive to $\delta \ll 1$. In order to avoid this 
problem we introduce the following new functions 
\begin{eqnarray}
  \Delta (x) & = & \frac{f(x)}{\cosh^2 (x)}, 
\label{sec4-10a}\\
  \psi (x) & = & \phi (x) + \ln \cosh (x) 
\label{sec4-10b}
\end{eqnarray}
where $x = r/a(0)^{1/2}$ is a dimensionless radius. 
It is necessary to note that when $f(x) = 1$ and $\phi (x) = 0$ 
we have the above-mentioned infinite flux tube. We have the following 
equations for these functions 
\begin{eqnarray}
  f'' - 4 f' \tanh (x) -2 f \left( 1 - 3 \tanh^2(x) \right) - 
  \frac{\left( f' - 2 f \tanh (x) \right)^2}{f} + &&  
\nonumber \\   
  \frac{a'}{a} \left( f' - 2 f \tanh (x) \right) + 
  \left( f' - 2 f \tanh (x) \right) 
  \left( \phi ' + \tanh (x) \right) + 
  \frac{q^2}{a^2 f} e^{-4 \phi}  = 0 , && 
\label{sec4-20a}\\
  a'' - 2 + a' \left( \phi ' + \tanh (x) \right) + 
  \frac{Q^2 f}{a} e^{2 \phi} = 0 , && 
\label{sec4-20b}\\
  \phi '' + \frac{1}{\cosh ^2(x)} + \left( \phi ' + \tanh (x) \right)^2 + 
  \frac{a'}{a}\left( \phi ' + \tanh (x) \right) - 
  \frac{Q^2 f}{2a^2} e^{2 \phi} = 0 && .
\label{sec4-20c}  
\end{eqnarray}
The corresponding solutions of these equations with the different $\delta$ 
are presented in Fig's \eqref{fig1}-\eqref{fig3}.
We see that (at least in the investigated area of $\delta$) at the point 
$r = \pm r_H$  the function $a(r_H) \approx 2 a_0$ (here $r_H$ is the point 
where $\Delta (r_H) = 0$ and $a_0 = a(r=0)$). The most interesting aspect
is that $a(r_H)$ does not grow as $r \rightarrow r_H$ and consequently 
\emph{the flux tube 
with $\delta \ll 1$ (in the region $|r| \leq r_H$) can be considered 
as a string-like object attached to two Universes.} For brevity we will 
call such objects \textit{threads}. 
\par 
It shows us that the necessary conditions 
for the consideration of our solution as a string-like object is definitely 
satysfied in the region $|r| \leq r_H$ and outside of this region 
the cross section will increase and consequently such approach will not 
be correct. For example, it is easy to see for the special case 
when $Q = 0$ \cite{dzhsin}. In this case there is the exact solution 
\begin{eqnarray}
a & = & r^{2}_{0} + r^{2},
\label{sec3-9a}\\
\Delta & = & \frac{q}{2r_{0} q}
\frac{r^{2}_{0} - r^{2}}{r^{2}_{0} + r^{2}},
\label{sec3-9b}\\
\psi &=& 0
\label{sec-9c}\\
\omega & = & \frac{4r_{0}}{q}
\frac{r}{r^{2}_{0} - r^{2}}
\label{sec3-9d}
\end{eqnarray}aAnd immediately we see that $a \approx r^2$ 
(at $a \rightarrow \infty$) and $g_{tt} \approx -1 + \mathcal O (1/r^2)$, 
the same is for the others cases with $Q < q$. It is interesting to note 
that here we have the change of the 4D metric signature \cite{dzhsch}. 
\par 
We can also insert this part of solution between two 
Reissner - Nordst\"om black holes \cite{dzh2} and join the flux tube 
with each black hole on the event horizon. For both cases for us is 
interesting only the part $|r| \leq r_H$ of the solution because here 
the transversal linear sizes are much less the longitudinal 
charachteristic length. 
\par 
Here we would like to investigate the behavior of all functions close to 
$r = r_H$. We propose the following approximate form of the metric 
\begin{eqnarray}
  \Delta (r) & = & \Delta_1 \left( r_H - r \right) + \cdots ,
\label{sec3-10a}\\
  a(r) & = & a_1 + a_2 \left( r_H - r \right) + \cdots ,
\label{sec3-10b}\\
  \psi (r) & = & \psi_1 + \psi_2 \left( r_H - r \right) + \cdots 
\label{sec3-10c}  
\end{eqnarray}
From Eq. \eqref{sec1-3} we see that 
\begin{equation}
  a_1 = \frac{q}{\Delta_1} e^{-2\psi_1}
\label{sec3-20}
\end{equation}
Now we would like to show that at this point we have 
$ds^2 = 0$ (where $d\chi = dr = d\theta = d\varphi = 0$ and  
$G_{55}(r = r_H) = 0$). The ($tt$) component of the metric is 
\begin{equation}
  G_{tt} (r) = \frac{1}{\Delta (r)} - R_1^2 \Delta (r)
  e^{2\psi (r)} \omega^2(r)
\label{sec3-30}
\end{equation}
According to Eq's \eqref{sec1-7e}, \eqref{sec3-10a}-\eqref{sec3-10c} 
we see that 
\begin{equation}
  \omega '(r) \approx \frac{q e^{-3\psi_1}}{R_1 a_1 \Delta_1^2 
  \left( r - r_H \right)^2}
\label{sec3-50}
\end{equation}
consequently
\begin{equation}
  \omega (r) \approx \frac{q}{R a_1 \left( r - r_H \right)} 
  e^{-3\psi_1}
\label{sec3-51}
\end{equation}
and 
\begin{equation}
  G_{tt} (r) \approx \frac{1}{\Delta_1 \left( r - r_H \right)^2}
  \left[
  1 - \frac{q^2 e^{-4\psi_1}}{a_1^2 \Delta_1^2}
  \right]
\label{sec3-60}
\end{equation}
from Eq. \eqref{sec3-20} we see that $G_{tt}(r_H) = 0$ and consequently 
$ds^2 = 0$. The same is true for $r = - r_H$.

\section{Gravitational waves on the thread.}

In this section we would like to consider the propagation of 
small perturbations (gravitational waves) of the 5D metric along 
the thread. It is not too hard to show \cite{landau} that the equations 
for these perturbations are 
\begin{eqnarray}
  h^C_{A;B;C} + h^C_{B;A;C} - {h_{AB}}^{;C}_{;C} - h_{;A;B} = 0 ,
\label{waves-10a}\\
  h_{AB} = G_{AB} - G^{(0)}_{AB} 
  \qquad \text{and} \qquad
  h = h^A_A
\label{waves-10b}
\end{eqnarray}
where $G^{(0)}_{AB}$ is the background metric \eqref{sec1-1} and 
$G_{AB}$ is the metric with the perturbations. Eq. \eqref{waves-10a} 
can be simplified in the case of waves with large frequency 
\begin{equation}
  {h_{AB}}^{;C}_{;C} = 0
\label{waves-20}
\end{equation}
with gauging 
\begin{equation}
  \left( h^B_A - \frac{1}{2} \delta^B_A h \right)_{;B} = 0 .
\label{waves-30}
\end{equation}
Let us introduce coordinates $\mathcal{X^A} = h_{AB}$, 
$\mathcal{A} = (A,B) = 0,1,2,\cdots ,15$. As mentioned above our thread 
approximately looks like a 1-dimensional object with a cross section 
of order of Planck scale $l^2_{Pl}$ and a classical longitudinal length.
This leads to the possibility that the gravitational waves depend
only on the time $t$ and the longitudinal coordinate $r$, \textit{i.e.} 
$h_{AB} = h_{AB}(t,r)$. Thus Eq. \eqref{waves-20} has the following 
form 
\begin{equation}
  \left( \mathcal{X^A}\right)^{;C}_{;C} = 0 
\label{vawes50}
\end{equation}
or 
\begin{equation}
  \left( \mathcal{X^A}\right)^{;a}_{;a} + 
  \left( \mathcal{X^A}\right)^{;m}_{;m} = 0
\label{waves60}
\end{equation}
where $a = t,r$ and $m = \chi ,\theta , \varphi$. We see that the second 
term does not contain derivatives. 
\par 
It is interesting to compare these equations for the thread with the 
classical string equations
\begin{equation}
  \Box X^\mu = \left( 
  \frac{\partial^2}{\partial\sigma^2} - 
  \frac{\partial^2}{\partial\tau^2}
  \right)X^\mu = 0
\label{waves70}
\end{equation}
where $X^\mu$ are the coordinates of the string and $\sigma ,\tau$ 
are the parameters on the sheet. We see that the difference is connected 
with the second term $( \mathcal{X^A})^{;m}_{;m}$. This difference 
between the thread and the string equations has the origin in the inner 
structure : the thread has a cross section ($\approx l^2_{Pl}$) but 
the string is a pure 1-dimensional object. Nevertheless, the gravitational 
objects on the thread can have propagating solutions.

\section{Discussion}

\begin{figure}
\begin{center}
\fbox{
\includegraphics[height=5cm,width=6cm]{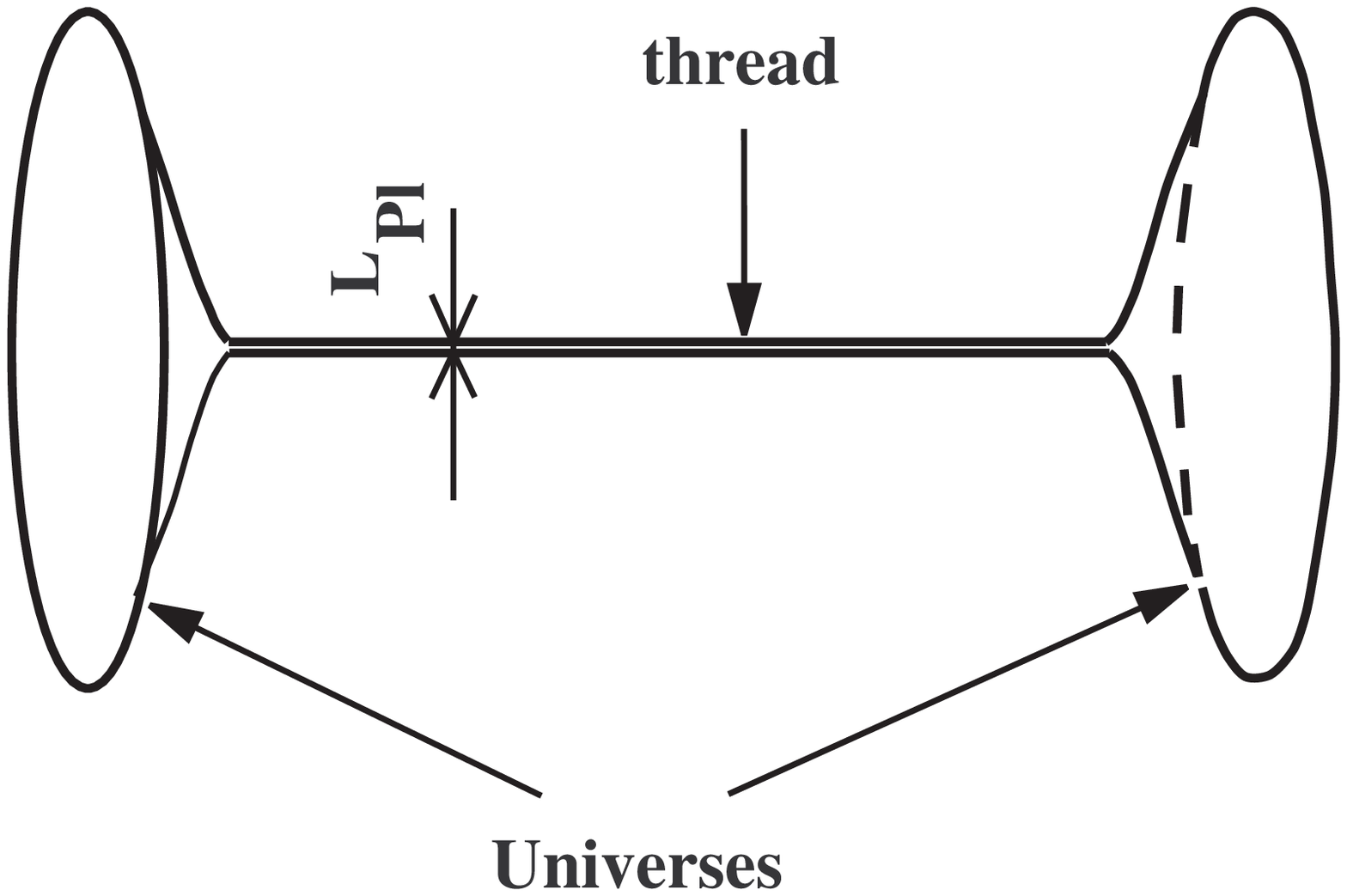}}
\caption{The long flux tube solution of the 5D Kaluza - Klein theory 
with (\textit{electric charge}) $\approx$ 
\textit{(magnetic charge)} can be considered as 
a thread with a Planckian cross section and classical 
longitudinal size attached to two Universes. 
Such a picture is very close to 
Guendelman's idea \cite{gundel} about the possibility of connecting 
a region where some dimensions are compact to another region where 
compactification does not exist.}
\label{fig5}
\end{center}
\end{figure}

Thus 5D Kaluza-Klein theory has a very interesting solutions which 
can be considered as a thread between two Universes 
(see, Fig. \eqref{fig5}). The cross section of such objects can be very 
small $\approx l_{Pl}$, but the longitudinal length can take any value. 
This remark gives us grounds to say that we have an unique object : 
in the transverse direction it is a quantum object, but in the 
longitudinal direction a classical one. It gives us the unique 
possibility to investigate quantum gravity only in one dimension. 
For example, a 1-dimensional spacetime foam on the thread and so on. 
\par 
Our thread presented on Fig. \eqref{fig5} looks like a string attached 
to two $D$-branes with the ends acting as sources of electric/magnetic 
charges in the external Universes. Nevertheless there is the 
difference : strings are 
located in some external spacetime but our threads do not need such 
space. Nevertheless, it is well known that any space with Lorentzian metric 
can be embedded into some multidimensional spacetime. 
\par 
Our investigation of the small perturbations propagating on the thread 
show us that the corresponding equations for the gravitational waves 
are sufficiently close to the classical string equations. Physically 
this difference appears because the thread has an inner transversal
structure whereas the string is a pure 1-dimensional object. 
\par 
Finally, we see in Kaluza - Klein gravity can exist 1-dimensional objects 
and the corresponding equations for the gravitational waves propagating 
on these objects are similar to the corresponding equations. 
Our investigation shows us that these 1-dimensional objects 
(threads in our notations) have an inner structure in the spirit of 
Einstein's idea that matter should be constructed from ``Nothing''. 

\section{Acknowledgment}
I am very grateful for Doug Singleton for the comments and fruitful 
discussion also DAAD for the financial support and Prof. Hagen Kleinert 
for the invitation for the research. 
Partially this work was supported by ISTC grant KR-677 and Alexander von 
Humboldt Foundation.

\end{document}